\documentclass[aps,onecolumn,superscriptaddress,groupedaddress]{revtex4}

\usepackage{amssymb}
\usepackage{color,graphicx}
\usepackage{amsmath}
\usepackage{amsbsy}
\usepackage{amsthm}
\usepackage{bbm}
\usepackage{bm}
\usepackage{epsfig}
\usepackage{lscape}
\usepackage{float}
\usepackage{subfigure}
\usepackage{dsfont}

\newcommand{\D}{\text{d}}
\newcommand{\I}{\imath}
\newcommand{\K}{\tilde{\mathbf{k}}}
\newcommand{\kk}{\mathbf{k}}
\newcommand{\com}[2]{\left[ {#1},{#2} \right]}

\begin{document}
\title{Testing Linearity of Quantum Theory with a Thermometer}

\author{M. Bahrami}
\email{mbahram@calstatela.edu}
\affiliation{Department of Chemistry, California State University, Los Angeles}

\begin{abstract}
Collapse models postulate that space is filled with a collapse noise field, inducing quantum Brownian motions which are dominant during the measurement, thus causing collapse of the wave function. An important manifestation of collapse noise field, if any, is thermal energy generation, thus disturbing the temperature profile of a system. The experimental investigation of collapse-driven heating effect has provided, so far, the most promising test of collapse models against standard quantum theory. In this paper, we calculate the collapse-driven heat generation for a three-dimensional multi-atomic Bravais lattice, by solving stochastic Heisenberg equations. We perform our calculation for the mass-proportional Continuous Spontaneous Localization collapse model with non-white noise. We obtain the temperature distribution of a sphere under stationary-state and adiabatic surface conditions. However, the exact quantification of effect highly depends on the value of cutoff in the collapse noise spectrum. 
\end{abstract}

\maketitle

\section{ Introduction} 
Recently, there has been an increasing interest in testing the linearity of quantum theory in unprecedented scales~\cite{exp_MW,exp_MW2,exp_MW3,dirk}. In particular, falsifying/verifying collapse models~\cite{GRW,CSL,collapse_review1,collapse_review2,adler} has received much theoretical and experimental attention~\cite{bassi,nimmrichter,diosi,goldwater,adler2,Vinante,Vinante2}. Collapse models postulate that the space is filled with a universal noise field that induces a quantum Brownian motion on the top of linear quantum dynamics. The collapse-driven Brownian corrections are dominant during the measurement process, thus causing the collapse of the wave function randomly with correct probabilities. The coupling with the collapse noise field has other important implications, e.g., it heats up the system, thus increasing its temperature~\cite{bassi,Vinante,Vinante2,adler2,adler3,diosi}. This unconventional manifestation has provided so far the most promising experimental falsification/verification of collapse models~\cite{Vinante,Vinante2}. Therefore, a more detailed calculation of the collapse-induced heating effect and how it disturbs the temperature profile of a system are needed. Here we calculate the rate of collapse-induced heat generation for a solid system using the quantum theory of solid state (e.g., see~\cite{Callaway}). We use the mass-proportional Continuous Spontaneous Localization collapse model since non-mass proportional models have been refuted experimentally~\cite{Piscicchia}. We also consider the non-white noise limit, because new results contradict the white-noise limit~\cite{Piscicchia}. However, as a practice, we will provide calculations for the white-noise because, first, to compare our results with the available ones in the literature, and second, to use some of them for the non-white case. 

The structure of this paper is as follow. First, we will introduce a linear stochastic Hamiltonian that mimics dynamical equation of collapse models in the white and nonwhite limits. We shall solve the stochastic Heisenberg equations of motion for the phononic annihilation and creation operator. After proper stochastic averaging, the rate of collapse-driven heat generation  will be calculated for the mono- and multi-atomic crystal under white and nonwhite noise limits. Finally, by solving heat diffusion equation under stationary-state and adiabatic boundary conditions, the temperature profile of a sphere will be obtained. We shall compare this result with the available experimental data. 

The reader should be noticed that in this paper, what we call as the density matrix is obtained after the stochastic averaging and it has no random terms (for more detail, see Sec.\ref{cumulantS} in the supplementary information). It should be distinguished from the density matrix in the Continuous Quantum Measurement literature which is still random (e.g., see~\cite{Bah-Juan} and references therein).

\section{Dynamical equations of collapse models} 
For the mass proportional Continuous Spontaneous Localization collapse model with a white noise, the stochastic-averaged dynamics of the density matrix reads as:
\begin{equation}
\label{cslmasteq2}
\frac{\D\hat{\rho}}{\D t}=-\frac{i}{\hbar}\left[\hat{H}_0,\hat{\rho}\right]-\frac12\gamma\int \frac{\D^3\K}{(2\pi)^3}\,e^{-r_c^2\K^2} 
\left[\hat{L}^\dagger(\K),\left[\hat{L}(\K),\hat{\rho}\right]\right],
\end{equation}
where the second term is the contribution of the collapse noise field, $\gamma=8\pi^{3/2}\lambda\,r_C^3$, and
\begin{equation}\label{F-density}
\hat{L}(\K)=\sum_{n=1}^\mathcal{N}\,\frac{M_n}{m_0}\,\exp[-\I\K\cdot\hat{\mathbf{x}}_n],
\end{equation}
is the Fourier transform of the number density operator where $\hat{\mathbf{x}}_n$ is the position operator of $n$-th atom, $M_n$ is the mass of $n$-th atom, $m_0=1\,$amu is the mass of one nucleon, $\mathcal{N}$ is the total number of atoms and $\I^2=-1$. $\lambda$ and $r_c$ are two phenomenological parameters defining the statistical properties of the collapse noise (see Eq.\eqref{Correlation}) with dimensions $[\lambda]=\text{s}^{-1}$ and $[r_C]=\text{m}$. In collapse models, the contribution of electrons to the collapse dynamics is negligible for all practical purposes and only nucleons are important. Note that $\hat{L}^\dagger(\K)=\hat{L}(-\K)$.
The dynamical equation in Eq.\eqref{cslmasteq2} can be reproduced by a linear stochastic Hamiltonian as follows:
\begin{equation}\label{H+V}
\hat{H}=\hat{H}_0+\hat{V}(t),
\end{equation}
with $\hat{H}_0$ the total Hamiltonian of the system and $\hat{V}(t)$ a stochastic interaction potential as:
\begin{equation}\label{V_t}
\hat{V}(t)=-\hbar\int\frac{\D^3 \K}{(2\pi)^3}\,e^{-r_c^2\K^2/2}\,\xi(t,\K)\,\hat{L}(\K),
\end{equation}
where $\xi(t,\K)=\D W(t,\K)/\D t$ is a complex Gaussian noise field where $W(t,\K)$ is a Wiener process. The statistical properties of the white noise field $\xi(t,\K)$ are:
\begin{equation}\label{Correlation}
\mathbb{E}(\xi(t,\K))=0;
~~~~
\mathbb{E}(\xi(t,\K)\xi(t',\K'))=(2\pi)^3\gamma\,\delta(t-t')\,\delta(\K+\K'),
\end{equation}
with $\mathbb{E}(\cdot)$ the stochastic-averaged value.
Note that $\xi^*(t,\K)=\xi(t,-\K)$.
For the non-white noise, one gets:
\begin{equation}\label{Correlation2}
\mathbb{E}(\xi(t,\K)\xi(t',\K'))=(2\pi)^3\,\delta(\K+\K')\,f(t-t');
~~~~~~~~~~~~
f(t)=\frac{1}{2\pi}\int_{-\infty}^{+\infty}\D\omega
\,\gamma(\omega)\,e^{-i\omega(t-t')},
\end{equation}
where $\gamma(\omega)$ is the collapse noise \textit{spectrum} (also called collapse noise \textit{spectral density}). For the white noise, $\gamma(\omega)$ is flat, i.e. $\gamma(\omega)=\gamma$.
For the stochastic Hamiltonian in Eq.\eqref{H+V} with nonwhite noise, the evolution of the density matrix will take a more complicated form. When the evolution of the density matrix is dominated by the second time-ordered cumulant (see Sec.\ref{cumulantS} in the supplementary information), the evolution reads as:
\begin{equation}
\label{mastereq-nonwhite}
\frac{\D\hat{\rho}}{\D t}=-\frac{i}{\hbar}\left[\hat{H}_0,\hat{\rho}\right]-\int \frac{\D^3\K}{(2\pi)^3}\,e^{-r_c^2\K^2} \int_0^t\D t'\,f(t-t')
\left[\hat{L}^\dagger(\K),\left[e^{\frac{i}{\hbar}\hat{H}_0(t'-t)}\hat{L}(\K)e^{-\frac{i}{\hbar}\hat{H}_0(t'-t)},\hat{\rho}\right]\right].
\end{equation}
Setting $f(t)=\gamma\delta(t)$ which is the white-noise limit, one recovers Eq.\eqref{cslmasteq2}.

In order to calculate any physical property of a system, we shall work with the stochastic Hamiltonian in Eq.\eqref{H+V} where the random potential is given in Eq.\eqref{V_t}. We will consider both white and nonwhite cases. All observable effects should be obtained after proper stochastic averaging. In particular, one should be careful if equations of motion are in Stratonovich or It\^o formalisms (see our explanation before Eq.\eqref{a(t)} or Sec.\ref{sto-switch} in the supplementary information).

\section{Three-dimensional Bravais lattice} 
We consider a three-dimensional crystal. We shall use labeling of atoms in a primitive cell which is very common in solid state textbooks (e.g. see~\cite[Sec.2.1]{Callaway}). 
A primitive cell is the parallelepiped defined by primitive vectors $\mathbf{a}_1$, $\mathbf{a}_2$ and $\mathbf{a}_3$. For example, for Copper with the face-centered cubic (fcc) crystal structure, $\mathbf{a}_1=a(1/2,1/2,0)$, $\mathbf{a}_2=a(0,1/2,1/2)$, and $\mathbf{a}_3=a(1/2,0,1/2)$ with $a=3.61478$\AA\, the lattice parameter (i.e. the size of unit cell) at $T=298$K~\cite[Table 1]{Cu}. The $i$-th primitive cell locates on the $i$-th lattice site and its position vector is given by $\mathbf{R}_i$, which is usually called as the lattice vector of the $i$-th lattice site. Note that any lattice vector $\mathbf{R}_i$ can be expressed as $\mathbf{R}_i=\sum_{l=1}^3n_{il}\mathbf{a}_l$ where $n_{il}$ are integers. The atoms within each primitive cell are indexed by Greek letters ($\nu, \kappa$) that run from $1$ to $r$ with $r$ the number of atoms in a primitive cell. We shall reserve Greek letters $\alpha, \beta$ for three coordinate components of any vector ($\alpha, \beta=1,2,3$). Therefore, a set of basis vectors $\mathbf{d}_\nu$ is needed to denote the position of atoms within a primitive cell. $\mathbf{d}_\nu$ is the vector connecting the tip of $\mathbf{R}_i$ with the $\nu$-th atom. 
For example, copper crystal is a mono-atomic crystal, thus having only one atom in a primitive cell (i.e. $\nu=1$). Or, TeO$_2$ is a multi-atomic crystal, with 12 atoms in its primitive cells (4 Te atoms and 8 oxygen atoms)~\cite[Table A.2]{TeO2-primitive}, thus $\nu=1,2,\cdots,12$. The equilibrium position of the $\nu$-th atom within the $i$-th primitive cell in a crystal is given by $\mathbf{R}_i+\mathbf{d}_\nu$.
Therefore, the $\alpha$-th component of the instantaneous position of $\nu$-th atom within the $i$-th primitive cell in a crystal can be written as classical equilibrium position plus a quantum displacement:
\begin{equation}\label{x-nu}
\hat{\mathbf{x}}_{\alpha,i\nu}=\mathbf{R}_{\alpha_,i}+\mathbf{d}_{\alpha,\nu}+\hat{\mathbf{u}}_{\alpha,i\nu}.
\end{equation}
where quantum displacement operator of $\hat{\mathbf{u}}_{i\nu}$, can be written in terms of the annihilation and creation operators of crystal phonons~\cite[Eq.(2.3.23)]{Callaway}:
\begin{equation}\label{displacement}
\hat{\mathbf{u}}_{\alpha,i\nu}=\sum_{\kk s}
\left(\frac{\hbar}{2NM_\nu\omega_{\kk s}}\right)^{1/2}
\left(\bm{\epsilon}^{(s)}_{\alpha,\nu}(\kk)\,e^{i\kk\cdot\mathbf{R}_i}\,\hat{a}_{\kk s}+
\bm{\epsilon}^{(s)*}_{\alpha,\nu}(\kk)\,e^{-i\kk\cdot\mathbf{R}_i}\,\hat{a}^\dagger_{\kk s}
\right),
\end{equation}
while the total Hamiltonian of the crystal is
\begin{equation}\label{2ndQF-Hamiltonian}
\hat{H}_0=\sum_{\kk s}\hbar\omega_{\kk s}\left(
\hat{a}^\dagger_{\kk s}\hat{a}_{\kk s}+1/2
\right).
\end{equation}
Note that $\kk$ is confined to the first Brillouin zone.
Also, there are in total $3rN$ phonon modes where $N$ is the total number of primitive cells and $r$ is the number of atoms in each primitive cell. For the total number of atoms in the crystal, $\mathcal{N}$, we have: $\mathcal{N}=rN$. 
The annihilation and creation operators $\hat{a}_{\kk s}$ and $\hat{a}^\dagger_{\kk s}$ satisfy the bosonic commutation relations:
\begin{equation}\label{commutation}
[\hat{a}_{\kk s}\hat{a}^\dagger_{\kk's'}]=\delta_{\kk,\kk'}\,\delta_{s,s'};~~~~~~~~
[\hat{a}_{\kk s}\hat{a}_{\kk's'}]=[\hat{a}^\dagger_{\kk s}\hat{a}^\dagger_{\kk's'}]=0.
\end{equation}
The polarization vectors $\bm{\epsilon}^{(s)}_\nu(\kk)$ are normalized vectors that describe the direction in which $\nu$-th atom moves. They satisfy orthonormality and completeness relations~\cite[Eq.(2.1.16)]{Callaway}:
\begin{equation}\label{polarization}
\sum_{\nu=1}^r\sum_{\alpha=1}^3\bm{\epsilon}^{(s)*}_{\alpha,\nu}(\kk)\bm{\epsilon}^{(s')}_{\alpha,\nu}(\kk)=\delta_{s,s'};~~~~~~~~
\sum_{s=1}^{3r}\bm{\epsilon}^{(s)*}_{\alpha,\nu}(\kk)\bm{\epsilon}^{(s)}_{\beta,\kappa}(\kk)=\delta_{\nu,\kappa}\delta_{\alpha,\beta},
\end{equation}
with $\bm{\epsilon}^{(s)*}_{\alpha,\nu}(\kk)=\bm{\epsilon}^{(s)}_{\alpha,\nu}(-\kk)$ where $\bm{\epsilon}^{(s)}_{\alpha,\nu}$ denotes $\alpha$-th  Cartesian component of the polarization vector of $\nu$-th atom. The value of $\omega_{\mathbf{k},s}$ is determined through the phonon dispersion relation $\omega_{\mathbf{k},s}=\sqrt{\lambda_s(\kk)}$ with $\lambda_s(\kk)$ the eigenvalues of the lattice dynamical matrix which is a $3r\times3r$ matrix whose eigenvectors are polarization vectors (e.g., see~\cite[Sec.(2.1.1)]{Callaway} for more detail). There are three acoustical branches (only one longitudinal) for $\omega_{\mathbf{k},s}$ and $3r-3$ optical branches ($r-1$ longitudinal branches). A monatomic crystal (i.e. $r=1$) has three branches of normal modes and they are all acoustical branches (one longitudinal and two transverse).

\section{Stochastic potential $\hat{V}(t)$ in terms of phonon operators} 
Introducing Eqs.\eqref{x-nu} and \eqref{displacement} into Eq.\eqref{F-density} yields
\begin{equation}\label{F-number-density}
\hat{L}(\K)=\sum_{i\nu}\,\frac{M_\nu}{m_0}\,e^{-\I\K\cdot(\mathbf{R}_i+\mathbf{d}_\nu)}
\prod_{\kk s}\hat{D}(\eta_{i\nu,\kk s}(\K)).
\end{equation}
where $\hat{D}(\eta_{i\nu,\kk s}(\K))$ is the \textit{phonon displacement} operator for the phonon mode $\kk s$~\footnote{Please note that $\hat{D}(\eta_{i\nu,\kk s}(\K))$ is the \textit{phonon} displacement operator for the mode $\kk s$, which is a very common terminology in quantum optics (e.g., ~\cite[Sec.11.3]{mw} or~\cite[Sec.5.3]{Loudon}). One should avoid the confusion with $\hat{\mathbf{u}}_{i\nu}$ which is the \textit{atom} displacement operator.}:
\begin{equation}
\hat{D}(\eta_{i\nu,\kk s}(\K))=\exp[
\eta_{i\nu,\kk s}(\K)\hat{a}^\dagger_{\kk s}-\eta^*_{i\nu,\kk s}(\K)\hat{a}_{\kk s}
]
\end{equation}
with
\begin{equation}\label{alpha}
\eta_{i\nu,\kk s}(\K)=-\I\left(\frac{\hbar}{2NM_\nu\omega_{\kk s}}\right)^{1/2}\sum_{\alpha=1}^3\K_\alpha\bm{\epsilon}^{(s)*}_{\alpha,\nu}(\kk)\,e^{-\I\kk\cdot\mathbf{R}_i}.
\end{equation}
Accordingly, the stochastic interaction Hamiltonian $\hat{V}(t)$ in Eq.\eqref{V_t} can be written as:
\begin{equation}\label{V_t-2nd}
\hat{V}(t)=-\hbar\int\frac{\D^3 \K}{(2\pi)^3}\,e^{-r_c^2\K^2/2}\,\xi(t,\K)\,\sum_{i\nu}\,\frac{M_\nu}{m_0}\,e^{-i\K\cdot(\mathbf{R}_i+\mathbf{d}_\nu)}
\prod_{\kk s}\hat{D}(\eta_{i\nu;\kk s}(\K)).
\end{equation}

\section{Heisenberg equations of motion} 
From now on, all the operators should be understood in the Heisenberg picture. Given the Hamiltonian in Eq.\eqref{H+V} together with Eqs.\eqref{2ndQF-Hamiltonian} and \eqref{V_t-2nd}, the Heisenberg equations of motion, in the Stratonovich formalism, yield:
\begin{eqnarray}
\frac{\D \hat{a}_{\kk s}}{\D t}=-\I\omega_{\kk s}\hat{a}_{\kk s}
+\I\int\frac{\D^3 \K}{(2\pi)^3}\,e^{-r_c^2\K^2/2}\,\xi(t,\K)\,\sum_{i\nu}\,\frac{M_\nu}{m_0}\,e^{-\I\K\cdot(\mathbf{R}_i+\mathbf{d}_\nu)}
\eta_{i\nu;\kk s}(\K)\hat{D}(\eta_{i\nu;\kk s}(\K))
\prod_{\kk' s'\neq \kk s}\hat{D}(\eta_{i\nu;\kk' s'}(\K))
\\
\frac{\D \hat{a}^\dagger_{\kk s}}{\D t}=\I\omega_{\kk s}\hat{a}^\dagger_{\kk s}
+\I\int\frac{\D^3 \K}{(2\pi)^3}\,e^{-r_c^2\K^2/2}\,\xi(t,\K)\,\sum_{i\nu}\,\frac{M_\nu}{m_0}\,e^{-\I\K\cdot(\mathbf{R}_i+\mathbf{d}_\nu)}
\eta^*_{i\nu;\kk s}(\K)\hat{D}(\eta_{i\nu;\kk s}(\K))
\prod_{\kk' s'\neq \kk s}\hat{D}(\eta_{i\nu;\kk' s'}(\K))
\end{eqnarray}
where we used the commutation relations $[\hat{D}(\eta_{i\nu;\kk' s'}(\K)),\hat{a}_{\kk s}]=-\eta_{i\nu;\kk s}(\K)\hat{D}(\eta_{i\nu;\kk s}(\K))\,\delta_{\kk,\kk'}\delta_{s,s'}$ and $[\hat{D}(\eta_{i\nu;\kk' s'}(\K)),\hat{a}^\dagger_{\kk s}]=-\eta^*_{i\nu;\kk s}(\K)\hat{D}(\eta_{i\nu;\kk s}(\K))\,\delta_{\kk,\kk'}\delta_{s,s'}$.
Since $|\eta_{i\nu;\kk s}(\K)|$ is proportional to $1/\sqrt{N}$ (see Eq.\eqref{alpha} where $N$ is the total number of primitive cells), for a large enough solid, we can safely assume that $|\eta_{i\nu;\kk s}(\K)|\ll1$. 
Accordingly, approximating the phonon displacement operator by $\hat{D}(\eta_{i\nu;\kk s}(\K))\approx\mathds{1}$, leads to:
\begin{eqnarray}
\frac{\D \hat{a}_{\kk s}}{\D t}&\approx&-\I\omega_{\kk s}\hat{a}_{\kk s}
+\I\int\frac{\D^3 \K}{(2\pi)^3}\,e^{-r_c^2\K^2/2}\,\xi(t,\K)\,\sum_{i\nu}\,\frac{M_\nu}{m_0}\,e^{-\I\K\cdot(\mathbf{R}_i+\mathbf{d}_\nu)}
\eta_{i\nu;\kk s}(\K)
\\
\frac{\D \hat{a}^\dagger_{\kk s}}{\D t}&\approx& \I\omega_{\kk s}\hat{a}^\dagger_{\kk s}
+\I\int\frac{\D^3 \K}{(2\pi)^3}\,e^{-r_c^2\K^2/2}\,\xi(t,\K)\,\sum_{i\nu}\,\frac{M_\nu}{m_0}\,e^{-\I\K\cdot(\mathbf{R}_i+\mathbf{d}_\nu)}
\eta^*_{i\nu;\kk s}(\K)
\end{eqnarray}
Note that above equations are the same in the Stratonovich and Ito formalisms because the terms containing random field (2nd terms in above equations) are independent from $\hat{a}_{\kk s}$ and $\hat{a}^\dagger_{\kk s}$~\footnote{In other word, the matrix $B^i(t)$ in Eq.\eqref{B-matrix} is independent from the random process $X_t$. Therefore, the second line in Eq.\eqref{Ito} will be zero and Eqs.\eqref{Stratonovich} and Eq.\eqref{Ito} will coincide.}. 
The solutions of above equations read as:
\begin{eqnarray}\label{a(t)}
\hat{a}_{\kk s}(t)&=&e^{-\I\omega_{\kk s}t}\hat{a}_{\kk s}+\I \int\frac{\D^3 \K}{(2\pi)^3}\,e^{-r_c^2\K^2/2}\,\int_0^t\D t'e^{-\I\omega_{\kk s}(t-t')}\xi(t',\K)\,\sum_{i\nu}\,\frac{M_\nu}{m_0}\,e^{-\I\K\cdot(\mathbf{R}_i+\mathbf{d}_\nu)}
\eta_{i\nu;\kk s}(\K)
\\\label{a*(t)}
\hat{a}^\dagger_{\kk s}(t)&=& e^{\I\omega_{\kk s}t}\hat{a}^\dagger_{\kk s}
+\I\int\frac{\D^3 \K}{(2\pi)^3}\,e^{-r_c^2\K^2/2}\,\int_0^t\D t'e^{\I\omega_{\kk s}(t-t')}\xi(t',\K)\,\sum_{i\nu}\,\frac{M_\nu}{m_0}\,e^{-\I\K\cdot(\mathbf{R}_i+\mathbf{d}_\nu)}
\eta^*_{i\nu;\kk s}(\K).
\end{eqnarray}
Using above equations, the stochastic-averaged Hamiltonian will be given as:
\begin{equation}\label{H-sto-ave}
\mathbb{E}(\hat{H}(t))=\sum_{\kk s}\, \hbar\omega_{\kk s}\mathbb{E}\left(\hat{a}^\dagger_{\kk s}(t)\hat{a}_{\kk s}(t)\right)
\end{equation}

\subsection{White noise field} 
Combining the white-noise correlations in Eq.\eqref{Correlation} together with Eq.\eqref{H-sto-ave} yields:
\begin{equation}
\mathbb{E}(\hat{H}(t))=
\hat{H}_0
-t\gamma\hbar\int\frac{\D^3 \K}{(2\pi)^3}\,e^{-r_c^2\K^2}\,
\sum_{i\nu,j\kappa}\frac{M_\nu M_\kappa}{m^2_0}\,e^{-\I\K\cdot(\mathbf{R}_i+\mathbf{d}_\nu)}\,e^{\I\K\cdot(\mathbf{R}_j+\mathbf{d}_\kappa)}
\sum_{\kk s}\omega_{\kk s}\eta^*_{i\nu;\kk s}(\K)\eta_{j\kappa;\kk s}(-\K)
\end{equation}
Introducing Eq.\eqref{alpha} into above equation gives:
\begin{eqnarray}
\mathbb{E}(\hat{H}(t))&=&
\hat{H}_0
+t\gamma\hbar\int\frac{\D^3 \K}{(2\pi)^3}\,e^{-r_c^2\K^2}\,
\sum_{i\nu,j\kappa}\frac{M_\nu M_\kappa}{m^2_0}\,e^{-\I\K\cdot(\mathbf{R}_i+\mathbf{d}_\nu)}\,e^{\I\K\cdot(\mathbf{R}_j+\mathbf{d}_\kappa)}
\times\\\nonumber&&~~~~~~~~~~~~~~~~~~~~~~~~~~~~~~~
\sum_{\kk }\frac{\hbar\,e^{\I \kk\cdot(\mathbf{R}_j-\mathbf{R}_i)}}{2N\sqrt{M_\nu M_\kappa}}
\sum_{\alpha,\beta=1}^{3}\K_\alpha\K_\beta\sum_s\bm{\epsilon}^{(s)*}_{\alpha,\nu}(\kk)\,\bm{\epsilon}^{(s)}_{\beta,\kappa}(\kk)
\\
&=&
\hat{H}_0
+\frac{\hbar^2 t\gamma \sum_\nu M_\nu}{2m^2_0}\int\frac{\D^3 \K}{(2\pi)^3}\,\K^2\,e^{-r_c^2\K^2}\,
\sum_{i,j}\,e^{-\I\K\cdot(\mathbf{R}_i-\mathbf{R}_j)}
\sum_{\kk}e^{\I\kk\cdot(\mathbf{R}_j-\mathbf{R}_i)}/N
\end{eqnarray}
where we used Eq.\eqref{polarization} to obtain the last line. By introducing the summation relation~\cite[Eq.(A.7)]{Callaway}:
$\sum_{\kk}e^{\I\kk\cdot(\mathbf{R}_j-\mathbf{R}_i)}/N=\delta_{i,j}$ into above equation, the stochastic-averaged Hamiltonian becomes:
\begin{eqnarray}
\mathbb{E}(\hat{H}(t))&=&
\hat{H}_0
+t\frac{\hbar^2 \gamma N\sum_\nu M_\nu}{2m^2_0}\int\frac{\D^3 \K}{(2\pi)^3}\,\K^2\,e^{-r_c^2\K^2}\,
\\\label{Heat-white}
&=&\hat{H}_0+t\frac{3\hbar^2 \lambda\mathcal{M}}{4m^2_0 r_c^2},
\end{eqnarray}
with $\mathcal{M}= N\sum_\nu M_\nu$ the total mass of the crystal. Note that in order to derive Eq.\eqref{Heat-white} the only approximation we used was when we obtained Eqs.\eqref{a(t)} and \eqref{a*(t)} by expanding the phonon displacement operators only to the first term. 

\subsection{Non-white noise field} 
Using the correlation of nonwhite noise in Eq.\eqref{Correlation2} together with Eq.\eqref{H-sto-ave}, the stochastic-averaged Hamiltonian reads as:
\begin{eqnarray}\label{H-nonwhite0}
\mathbb{E}(\hat{H}(t))&=&
\hat{H}_0
+\frac{\hbar^2}{2V_{BZ}}\sum_{\nu,\kappa}\frac{\sqrt{M_\nu M_\kappa}}{m^2_0}\,\int\frac{\D^3 \K}{(2\pi)^3}\,e^{-r_c^2\K^2}\,e^{-\I\K\cdot(\mathbf{d}_\nu-\mathbf{d}_\kappa)}
\times\\\nonumber&&
\int_{BZ}\frac{\D^3\kk}{(2\pi)^3}\,\sum_{i,j}e^{\I (\K+\kk)\cdot(\mathbf{R}_j-\mathbf{R}_i)}\,
\sum_{s}\int_0^t\D t_1\int_0^t\D t_2e^{-\I\omega_{\kk s}(t_2-t_1)}f(t_2-t_1)\,\sum_{\alpha,\beta=1}^{3}\K_\alpha\K_\beta\bm{\epsilon}^{(s)*}_{\alpha,\nu}(\kk)\,\bm{\epsilon}^{(s)}_{\beta,\kappa}(\kk)
\end{eqnarray}
where $V_{BZ}=(2\pi)^3/|\mathbf{a}_1\cdot(\mathbf{a}_2\times\mathbf{a}_3)|$ is the volume of the first Brillouin zone. We have replaced the summation over $\kk$ by an integration: $\sum_{\kk}=(1/V_{BZ})\int\D^3\kk/(2\pi)^3$. 

For the time-dependent part in above equation, we have:
\begin{eqnarray}
\int_0^t\D t_1\int_0^t\D t_2e^{-\I\omega_{\kk s}(t_2-t_1)}f(t_2-t_1)&=&
\int_{-\infty}^{\infty}\frac{\D \omega}{2\pi}\gamma(\omega)\int_0^t\D t_1e^{i(\omega_{\kk s}-\omega)t_1}\int_0^t\D t_2e^{-i(\omega_{\kk s}-\omega)t_2}
\\&=&
\int_{-\infty}^{\infty}\frac{\D \omega}{2\pi}\gamma(\omega)\int_{-t/2}^{t/2}\D t_1e^{i(\omega_{\kk s}-\omega)t_1}\int_{-t/2}^{t/2}\D t_2e^{-i(\omega_{\kk s}-\omega)t_2}
\\&=&
\int_{-\infty}^{\infty}\frac{\D \omega}{2\pi}\gamma(\omega)\left(2\pi\delta^{(t)}(\omega_{\kk s}-\omega)\right)^2.
\end{eqnarray}
where $2\pi\delta^{(t)}(\omega)=\int_{-t/2}^{t/2}\D t'e^{-\I\omega t'}$. Since $\delta^{(t)}(\omega_{\kk s}-\omega)$ has its maximum at $\omega_{\kk s}=\omega$ and falls rapidly to zero elsewhere, to a first approximation one can replaced $\left(\delta^{(t)}(\omega_{\kk s}-\omega)\right)^2\approx\delta^{(t)}(\omega_{\kk s}-\omega)\delta^{(t)}(0)=(t/2\pi)\,\delta^{(t)}(\omega_{\kk s}-\omega)$. Also when $\omega_{\kk s}t\gg1$, one can approximate $\delta^{(t)}(\omega_{\kk s}-\omega)\approx\delta(\omega_{\kk s}-\omega)$ (see~\cite[p.443]{mw},~\cite[Sec.2.4]{Loudon} or~\cite[p.129]{gc}). Introducing these approximations into above equations yields:
\begin{eqnarray}
\int_0^t\D t_1\int_0^t\D t_2e^{-\I\omega_{\kk s}(t_2-t_1)}f(t_2-t_1)&\approx&
t\gamma(\omega_{\kk s}).
\end{eqnarray}
Putting above equation into Eq.\eqref{H-nonwhite0} leads to:
\begin{eqnarray}\label{H-nonwhite}
\mathbb{E}(\hat{H}(t))&=&
\hat{H}_0
+\frac{t\hbar^2}{2V_{BZ}}\sum_{\nu,\kappa}\frac{\sqrt{M_\nu M_\kappa}}{m^2_0}\,\int\frac{\D^3 \K}{(2\pi)^3}\,e^{-r_c^2\K^2}\,e^{-\I\K\cdot(\mathbf{d}_\nu-\mathbf{d}_\kappa)}
\times\\\nonumber&&~~~~~~~~~~~~~~~~~~~~~~~~
\int_{BZ}\frac{\D^3\kk}{(2\pi)^3}\,\sum_{i,j}e^{\I (\K+\kk)\cdot(\mathbf{R}_j-\mathbf{R}_i)}\,
\sum_{s}\gamma(\omega_{\kk s}) \,\sum_{\alpha,\beta=1}^{3}\K_\alpha\K_\beta\bm{\epsilon}^{(s)*}_{\alpha,\nu}(\kk)\,\bm{\epsilon}^{(s)}_{\beta,\kappa}(\kk)
\end{eqnarray}
Note that in Eq.\eqref{H-nonwhite} the $\K$-integration is over all space and while $\kk$ is confined inside the first Brillouin zone. Since the lattice constant is usually of the order $a\sim10^{-10}-10^{-9}\,$m, therefore $V_{BZ}\sim10^{29}-10^{33}\,\text{m}^{-3}$. The volume in which the most important $\K$ are inside, is $V_{c}\sim(2\pi/r_c)^3~\sim10^{23}\,\text{m}^{-3}$ ($r_c=10^{-7}\,$m). This implies that the important $\K$ are well-inside the first Brillouin zone. Therefore, we can safely use the following summation relation~\cite[Eq.(A.9b)]{Callaway}:
\begin{equation}
\sum_{i}e^{\I (\K+\kk)\cdot\mathbf{R}_i}=V_{BZ}\,\delta(\kk+\K)
\end{equation}
Introducing above equation into Eq.\eqref{H-nonwhite} together with $\bm{\epsilon}^{(s)}_{\kappa}(\kk)=\bm{\epsilon}^{(s)*}_{\kappa}(-\kk)$, and $\omega_{\kk s}=\omega_{-\kk s}$ yields
\begin{eqnarray}\label{Heat-nonwhite-M0}
\mathbb{E}(\hat{H}(t))&=&\hat{H}_0
+\frac{t\hbar^2\,N}{2m_0^2}\,\int\frac{\D^3 \kk}{(2\pi)^3}\,e^{-r_c^2\kk^2}\,
\sum_{s}\gamma(\omega_{\kk s})\,\left|\sum_{\nu}\sqrt{M_\nu}e^{-\I\kk\cdot\mathbf{d}_\nu}\,\sum_{\alpha=1}^{3}\kk_\alpha\bm{\epsilon}^{(s)}_{\alpha,\nu}(\kk)\right|^2.
\end{eqnarray}
Note that the $\kk$-integration is now over the entire reciprocal space. The error introduced by this extension is negligible because the exponential term $e^{-r_c^2\kk^2}$ will make sure that the important modes are well-inside the Brillouin zone. Also, if the polarization is perpendicular to $\kk$, it will not have any contribution, thus only longitudinal polarizations will survive in Eq.\eqref{Heat-nonwhite-M0}.
For the special case of a mono-atomic crystal (i.e. $\nu=1$), the polarization vectors are only acoustic: one longitudinal (LA), two transverse modes (TA$_{1,2}$). Therefore, Eq.\eqref{Heat-nonwhite-M0} reduces to:
\begin{eqnarray}\label{Heat-nonwhite-1}
\mathbb{E}(\hat{H}(t))&=&\hat{H}_0
+\frac{t\hbar^2 \mathcal{M}}{2m_0^2}\,\int\frac{\D^3 \kk}{(2\pi)^3}\,\kk^2\,e^{-r_c^2\kk^2}\,\gamma(\omega_{\kk \text{LA}}),
\end{eqnarray}
where $\mathcal{M}=N M$ is the total mass of the mono-atomic crystal. 

We can simplify Eq.\eqref{Heat-nonwhite-M0} more. Because of the exponential term $e^{-r_c^2\kk^2}$ in Eq.\eqref{Heat-nonwhite-M0}, we have $|\kk|\lesssim 1/r_c$; while $|\mathbf{d}_\nu|\leq a$ with $a\lesssim10^{-9}\,\text{m}$ the typical value of lattice constant. Since $r_c\sim10^{-7}\,$m, therefore one gets $r_c\gg a$ which leads to $e^{-\I\kk\cdot\mathbf{d}_\nu}\approx1$. In addition, for $r_c\gg a$, one can safely use the long-wave limit (see~\cite[Sec.4.3.3]{Lev} and references therein) as follows:
\begin{eqnarray}
\mathbb{E}(\hat{H}(t))&=&\hat{H}_0
+\frac{t\hbar^2\,N}{2m_0^2}\,\int\frac{\D^3 \kk}{(2\pi)^3}\,\,e^{-r_c^2\kk^2}\,
\sum_{s}\gamma(\omega_{\kk s})\,\left|\sum_{\alpha=1}^3\kk_\alpha\sum_{\nu}\sqrt{M_\nu}\,\bm{\epsilon}^{(s)}_{\alpha,\nu}(0)\right|^2,
\end{eqnarray}
where, to the leading orders in $a/r_c$, we replaced $\bm{\epsilon}^{(s)}_{\alpha,\nu}(\kk)\approx\bm{\epsilon}^{(s)}_{\alpha,\nu}(0)$. In the long-wave limit, the optical branches (denoted by $s_\text{O}$) obey~\cite[Eq.(4.106)]{Lev}:
$\sum_{\nu}\sqrt{M_\nu}\,\bm{\epsilon}^{(s_\text{O})}_{\alpha,\nu}(0)=0$; while, for the acoustic branches (denoted by $s_\text{A}$) we have~\cite[Eqs.(4.84,4.103)]{Lev}: $\left|\bm{\epsilon}^{(s_\text{A})}_{\nu}(0)/\sqrt{M_\nu}\right|=\left(\sum_\kappa M_\kappa\right)^{-1/2}$. Accordingly, to the leading order of $a/r_c$, the above equation is dominated only by the longitudinal acoustic branch:
\begin{eqnarray}\label{heat-multi-LA}
\mathbb{E}(\hat{H}(t))&=&\hat{H}_0
+\frac{t\hbar^2\,\mathcal{M}}{2m_0^2}\,\int\frac{\D^3 \kk}{(2\pi)^3}\,\kk^2\,e^{-r_c^2\kk^2}\,
\gamma(\omega_{\kk \text{LA}}),
\end{eqnarray}
where $\mathcal{M}=N \sum_\nu M_\nu$ is the total mass of the multi-atomic crystal. Note that Eq.\eqref{heat-multi-LA} is the same as the one for the mono-atomic crystal in Eq.\eqref{Heat-nonwhite-1}.

In the long-wave limit, the Debye model can be applied for the longitudinal acoustic branch as follows (e.g., see~\cite[Sec.4.3.5]{Lev} for more detail): 
\begin{equation}
\int\frac{\D^3\kk}{(2\pi)^3}\,f(\omega_{\kk \text{LA}})=\int_0^{\omega_{\text{D}}}\frac{\omega^2}{2\pi^2v_{\text{eff}}^3}f(\omega)\D\omega,
\end{equation}
with $\omega_{\text{D}}=v_{\text{eff}}(6\pi^2/V_0)^{1/3}$ the Debye frequency, $v_{\text{eff}}$ the effective sound velocity in the crystal, and $V_0$ the volume of primitive cell ($V_0\sim a^3$). 
Putting these results into Eqs.\eqref{Heat-nonwhite-1} and \eqref{heat-multi-LA} leads to:
\begin{eqnarray}\label{Heat-nonwhite-M}
\mathbb{E}(\hat{H}(t))&=&\hat{H}_0
+\frac{t\,\hbar^2\mathcal{M}}{4\pi^2m_0^2v_{\text{eff}}^5} 
\int_0^{\infty}\D\omega\,\omega^4\,e^{-\omega^2r_c^2/v_{\text{eff}}^2}\,\gamma(\omega),
\end{eqnarray}
where the error introduced by extending the interval of integration to infinity is negligible because $\omega_{\text{D}}\gg v_{\text{eff}}/r_c$ for $r_c\gg a$ (the exponential term $e^{-\omega^2r_c^2/v_{\text{eff}}^2}$ introduces a frequency cut-off $v_{\text{eff}}/r_c$).

\section{Concluding remarks} 
A very interesting manifestation of collapse models is a \textit{uniform} heat generation throughout a sample. The thermal equilibrium (i.e. a uniform temperature throughout) will be disturbed by the collapse-induced heat generation. Therefore, a temperature distribution (say, temperature profile or field) will be produced in the system, which can be used to falsify/verify collapse models against standard quantum predictions~\cite{adler2}.
The change in the temperature profile of the system can be obtained from the heat diffusion equation~\cite[Eq.(2.29)]{Frank}:
\begin{equation}
\vec{\nabla}\cdot(k\vec{\nabla}T)
+\dot{q}=\varrho c_p\frac{\partial T}{\partial t}
\end{equation}
where $\varrho$ is the mass density, $c_p$ is the specific heat capacity at constant pressure, $k$ is the thermal conductivity, and $\dot{q}$ is the rate of energy generation per unit volume. For the rate of heat generation, we have: 
\begin{equation}
\dot{q}=\frac{1}{V}\frac{\D}{\D t}\mathbb{E}(\hat{H}(t)),
\end{equation}
with $V$ the volume of the object.
Therefore, using Eq.\eqref{Heat-white} for the white noise, the rate of thermal energy generation per unit volume is given by:
\begin{eqnarray}\label{heat-rate-white}
\dot{q}_{\text{w}}=\frac{3\hbar^2 \lambda\,\varrho}{4m^2_0 r_c^2}
\end{eqnarray} 
where $\varrho$ is the mass density of crystal.
For a nonwhite noise field in Eq.\eqref{Heat-nonwhite-M}, the rate of thermal energy generation per unit volume is:
\begin{eqnarray}\label{heat-rate}
\dot{q}_{\text{nw}}=
\frac{\hbar^2\varrho}{4\pi^2m_0^2v_{\text{eff}}^5}
\int_0^{\infty}\D\omega\,\omega^4\,e^{-\omega^2r_c^2/v_{\text{eff}}^2}\,\gamma(\omega).
\end{eqnarray}
The collapse noise spectrum, $\gamma(\omega)$, can be simply written as the white noise strength multiplied by a frequency cut-off term as:
$\gamma(\omega)=(8\pi^{3/2}\lambda\,r_c^3)\,\theta(\Omega-\omega)$, 
with $\Omega$ the frequency cut-off and $\theta(\cdot)$ the Heaviside step function. 
Therefore, $\dot{q}_{\text{nw}}$ in two limiting cases obey:
$\dot{q}_{\text{nw}} = \dot{q}_{\text{w}}$ for $\Omega\gg v_{\text{eff}}/r_c$ and 
$\dot{q}_{\text{nw}} =\frac{8 (\Omega r_c/v_{\text{eff}})^5}{15\sqrt{\pi}}\,\dot{q}_{\text{w}}$ for $\Omega\ll v_{\text{eff}}/r_c$. As clear, for $\Omega\ll v_{\text{eff}}/r_c$, one gets $\dot{q}_{\text{nw}} \ll \dot{q}_{\text{w}}$. Therefore, the heating effect of a non-white collapse noise can be equal or smaller than that of a white noise, which depends on the value of frequency cut-off $\Omega$. 

We now apply the results in Eqs.(\ref{heat-rate-white}) and \eqref{heat-rate} for a one-dimensional (i.e. temperature gradient along only one coordinate), steady state case with adiabatic surfaces (e.g., see~\cite[Fig.C.2]{Frank}). Let us consider a spherical object with the radius $r_0$, which is initially prepared at a very low temperature $T_{\text{s}}$, and it will be in contact with the thermal bath $T_{\text{s}}$. Under the steady-state condition ($\partial T/\partial t=0$), the heat diffusion equation reduces to:
\begin{equation}
\frac{1}{r^2}\frac{\D}{\D r}\left(kr^2\frac{\D T}{\D r}\right)
+\dot{q}=0,
\end{equation}
with $r$ the distance from the center of sphere. The adiabatic boundary condition is~\cite[Eq.(2.33)]{Frank}:
\begin{equation}\label{boundary}
T(r_0)=T_{\text{s}};~~~~~~~~~~~~~~~~~~~
\left.\frac{\D T}{\D r}\right|_{r=0}=0.
\end{equation}
Since we are interested in cases where $T_{\text{s}}$ is very low ($T_{\text{s}}\ll1\,$K), one can approximate the thermal conductivity by $k=k_0T$, which implies:
\begin{equation}\label{head-D-Eq}
\frac{k_0}{r^2}\frac{\D}{\D r}\left(r^2T\frac{\D T}{\D r}\right)
+\dot{q}=0.
\end{equation}
Note that $\dot{q}$ is independent from $r$. Considering boundary conditions in Eq.\eqref{boundary}, the temperature profile $T(r)$ reads as:
\begin{eqnarray}\label{sphere}
T(r)=T_{\text{s}}\sqrt{1+\frac{\dot{q}}{3k_0T_{\text{s}}^2}\left(1-\frac{r^2}{r_0^2}\right)}\approx
T_{\text{s}}+\frac{\dot{q}}{6k_0T_{\text{s}}}\left(1-\frac{r^2}{r_0^2}\right).
\end{eqnarray}
where the approximation is valid if $\dot{q}\ll k_0T_{\text{s}}$, which is the case since the collapse-driven heating effect is tiny. It is interesting that the approximation in Eq.\eqref{sphere} leads to the same equation as that of a sphere with constant thermal conductivity $k=k_0T_{\text{s}}$ (see~\cite[Eq.(C.24)]{Frank}). Using Eq.\eqref{sphere}, the core temperature (the temperature at the center of object) will be
\begin{equation}\label{core-T}
T_c=T(r=0)=T_{\text{s}}+\frac{\dot{q}}{6k_0T_{\text{s}}}.
\end{equation}
Now we will use above equation to calculate the core temperature for two experimental cases. 

First, a piece of copper that is cooled down to $30\,$mK with $k_0=80-170\,\text{W}/(\text{m}\cdot\text{K}^2)$~\cite[Tabel 1]{NOSV-Cu} and $\varrho_{\text{Cu}} =8.90\times10^3\,\text{kg}/\text{m}^3$. 
For simplicity, we would assume it is a spherical object.
For the white noise limit, the core temperature reads as:
$T_c\approx T_{\text{s}}+200\lambda$ for $k_0=80\,\text{W}/(\text{m}\cdot\text{K}^2)$ and 
$T_c\approx T_{\text{s}}+100\lambda$ for $k_0=170\,\text{W}/(\text{m}\cdot\text{K}^2)$.
Considering the most promising value of $\lambda\sim10^{-8}\,\text{s}^{-1}$~\cite{adler}, one finds:
$T_c-T_{\text{s}}\sim10^{-6}\,$K, which is four order of magnitude smaller than the temperature achieved in CUORE experiment~\cite{NOSV-Cu,CUORE}. For a non-white collapse noise, the difference between core and surface temperature will be less or equal to $10^{-6}\,$K.

As another example, let us consider a $5\text{cm}\times5\text{cm}\times5\text{cm}$ cube of TeO$_2$ crystal of mass 750g which has been cooled down to 10mK in the CUORE experiment~\cite[Sec.2,1st par.]{CUORE}. For simplicity, let us replace the $5\text{cm}\times5\text{cm}\times5\text{cm}$ cube of TeO$_2$ crystal by a sphere of radius $r_0=3.1\,$cm (same mass, same number of primitive cells). 
Setting $k=3\,\text{W}/(\text{m}\cdot\text{K}^2)$ and assuming a white noise, the difference between the core and surface temperatures will be:
$T_c - T_{\text{s}}\approx 10^{4}\lambda$. For $\lambda\sim10^{-8}\,\text{s}^{-1}$, one gets: $T_c - T_{\text{s}}\sim 10^{-4}\text{K}$, which is two order of magnitude larger than that of copper in the previous example, but still two order of magnitude smaller than the lowest temperature reached in CUORE experiment~\cite{CUORE}. For this case, the temperature difference for a non-white collapse noise will be equal or smaller than $\sim 10^{-4}\text{K}$.

As clear, very dense materials with small thermal conductivity (see Eqs.\eqref{heat-rate-white} and \eqref{core-T}) seem to be best candidates to investigating the collapse heating effects. Therefore, nonconducting solids are more promising since they have smaller thermal conductivity than the conducting ones (e.g., see~\cite[Sec.2.2.1]{Frank}).
It is worth mentioning that the temperature difference between the center and the surface also depends on the shape (i.e. geometry). For example, a plane wall (i.e. a cuboid with a width much smaller that length and height) with the same characteristics as a sphere will have a temperature difference three times larger than that of sphere (e.g., see~\cite[Eqs(C.22-24)]{Frank}).

In conclusion, collapse models predict a uniform heat generation that disturbs the temperature profile of a system initially prepared in thermal equilibrium. Therefore, observing the change in the temperature profile is another way to test quantum theory against collapse models. The most promising candidates are dense solids with very low thermal conductivity. However, the results highly depend the value of cutoff in the collapse noise spectrum $\gamma(\omega)$.

\subsection{Final comment}
While preparing v1 of this paper, we became aware of a very recent work by Adler (v1 announced on arXiv on 01/01/18) in which Adler has calculated the collapse heat generation using a perturbative approach~\cite{adler3}. We would like to mention that our Eqs.\eqref{Heat-white} and (\ref{Heat-nonwhite-1},\ref{heat-multi-LA}) are the same as Eqs.15 and 17 in~\cite{adler3}. However, our calculation is not perturbative, and it is completely independent from the initial state of the system. In the 2nd version of our paper (v2 announced on arXiv on 01/14/18), we had used a wrong normalization factor for the polarization of multi-atomic crystals and, therefore, we got a wrong formula for the non-white case of a multi-atomic crystal. Adler correctly commented on that~\cite[Appendix]{adler3} (v3 announced on arXiv on 01/24/18) and we've revised our calculation accordingly in this version.  

\section*{Acknowledgment}
MB thanks Alireza Bafandeh of University of North Carolina at Charlotte (USA), and Mahdi Bahrami of MAPNA group (IRAN), for their valuable help on the heat transfer equations. MB also thanks Dr. Giulio Gasbarri of Trieste University (ITALY) for valuable discussions that we had on cumulant expansion techniques. MB would like to thank Prof. Steve Adler of Institute for Advanced Study at Princeton (USA), for his valuable comment on multi-atomic crystals, and also Prof. Lajos Di\'osi of HAS, Wigner Research Centre for Physics at Budapest (HUNGARY) for his comments on our references in the introduction.



\onecolumngrid

\setcounter{equation}{0}
\setcounter{section}{0}
\renewcommand{\theequation}{S.\arabic{equation}}
\renewcommand{\thesection}{S.\arabic{section}}

\section{Supplementary Material}
We briefly discuss the derivation of a master equation from a stochastic linear Schr\"odinger equation. We also review how to switch between Stratonovich and It\^o stochastic differential equations. The reader should notice that here we just summarize the important available results in the literature to the extend that we need them for our derivation in the main text. For further details, we refer the reader to references in the text.

\subsection{Time-ordered cumulant expansion method for the master equation}\label{cumulantS}
In this section we shall report important results in the literature on deriving a master equation from a stochastic linear Schr\"odinger equation. If we consider a Schr\"odinger equation with a random potential, then all observable results should be obtained after taking the stochastic average over the random field. Therefore, one can work at the level of master equation for the density matrix. The density matrix is given by $\hat{\rho}(t)=\mathbb{E}(|\psi(t)\rangle\langle\psi(t)|)$ with $\mathbb{E}(\cdot)$ denoting the ensemble-averaging over all possible realizations of the random field (say, stochastic averaging).
In the following, we will derive the master equation by using time-ordered cumulant technique~\cite{Kampen,Fox74,Fox75,Fox76}. In order to study the evolution of density matrix, it is more convenient, however equivalent, to work with the following equation in the interaction picture:
\begin{align}\label{evol-I}
\frac{\D}{\D t}(|\psi^{\text{\tiny I}}(t)\rangle\langle\psi^{\text{\tiny I}}(t)|)&=
\I\hat{\cal V}(t)\,|\psi^{\text{\tiny I}}(t)\rangle\langle\psi^{\text{\tiny I}}(t)|,
\end{align}
with $\hat{\cal V}(t)$ a super-operator such that $\hat{\cal V}(t)\,|\psi^{\text{\tiny I}}(t)\rangle\langle\psi^{\text{\tiny I}}(t)|=\left[\hat{V}^{\text{\tiny I}}(t),|\psi^{\text{\tiny I}}(t)\rangle\langle\psi^{\text{\tiny I}}(t)|\right]$ where $\hat{V}^{\text{\tiny I}}(t)$ is the random potential in the interaction picture (e.g., the interaction picture of potential in Eq.(4) in the main text).

Given Eq.\eqref{evol-I}, the stochastic unitary evolution operator that maps $|\psi^{\text{\tiny I}}(0)\rangle\langle\psi^{\text{\tiny I}}(0)|$ to $|\psi^{\text{\tiny I}}(t)\rangle\langle\psi^{\text{\tiny I}}(t)|$ is given by:
\begin{equation}\label{U}
\hat{\cal U}^{\text{\tiny I}}(t;t_0)=\left\lceil\exp\left(\I\int_{t_0}^t\D s\,\hat{\cal V}(s)\right)\right\rceil,
\end{equation}
where $\left\lceil\cdots\right\rceil$ denotes the time ordering. We set $t_0=0$ when the non-random state $|\psi(0)\rangle$ is prepared, and $\hat{\cal U}^{\text{\tiny I}}(0;0)=\hat{\mathds{1}}$ which is also non-random.
Therefore, one finds:
\begin{align}
|\psi^{\text{\tiny I}}(t)\rangle\langle\psi^{\text{\tiny I}}(t)|=
\hat{\cal U}^{\text{\tiny I}}(t;0)\,|\psi(0)\rangle\langle\psi(0)|.
\end{align}
Notice that in above equation, only the mapping $\hat{\cal U}^{\text{\tiny I}}(t;0)$ is stochastic while $|\psi(0)\rangle\langle\psi(0)|$ is a non-random term.
Accordingly, for the density operator we get:
\begin{align}\label{U-I}
\hat{\rho}^{\text{\tiny I}}(t)=
\mathbb{E}\left(|\psi^{\text{\tiny I}}(t)\rangle\langle\psi^{\text{\tiny I}}(t)|\right)=
\mathbb{E}\left(\hat{\cal U}^{\text{\tiny I}}(t;0)\right)\,|\psi(0)\rangle\langle\psi(0)|.
\end{align}
Given Eq.\eqref{U-I}, the evolution of the destiny matrix can be obtained once $\frac{\D}{\D t}\mathbb{E}\left(\hat{\cal U}^{\text{\tiny I}}(t;0)\right)$ is determined. Using the time-ordered cumulant technique~\cite{Kampen,Fox74,Fox75,Fox76}, the stochastic-average $\mathbb{E}\left(\hat{\cal U}^{\text{\tiny I}}(t;0)\right)$ is given by:
\begin{align}\label{U-cum}
\mathbb{E}\left(\hat{\cal U}^{\text{\tiny I}}(t;0)\right)
=\mathbb{E}\left(\left\lceil\exp\left(i\int_{0}^t\D s\,\hat{\cal V}(s)\right)\right\rceil\right)=
\left\lceil\exp\left(\sum_{n=1}^\infty\int_{0}^t\D s\,\hat{\cal G}^{(n)}(s)\right)\right\rceil
\end{align}
where $\hat{\cal G}^{(n)}(s)$ is:
\begin{align}\label{G-n}
\hat{\cal G}^{(n)}(t)=i^n\,\int_0^t\D t_2\int_0^{t_2}\D t_3\cdots\int_0^{t_{n-1}}\D t_n\,
\mathbb{E}^{(n)}_c\left(
\hat{\cal V}(t)\,\hat{\cal V}(t_2)\,\hat{\cal V}(t_3)\cdots\hat{\cal V}(t_n)
\right)
\end{align}
with $\mathbb{E}^{(n)}_c\left(\cdot\right)$ the $n$-th order time-ordered cumulant stochastic-average.
The general expression for $\mathbb{E}^{(n)}_c\left(\cdot\right)$ in terms of $\mathbb{E}\left(\cdot\right)$ is provided by van Kampen~\cite{Kampen} and Fox~\cite{Fox74,Fox75,Fox76} (in particular, see Ref.~\cite{Fox75} for a detailed derivation). Briefly, the time-ordered cumulant and thus $\hat{\cal G}^{(n)}(t)$ can be computed by introducing a parameter $\lambda$: $\mathbb{E}\left(\left\lceil\exp\left(i\lambda\int_{0}^t\D s\,\hat{\cal V}(s)\right)\right\rceil\right)=\left\lceil\exp\left(\sum_{n=1}^\infty\int_{0}^t\D s\,\lambda^n\hat{\cal G}^{(n)}(s)\right)\right\rceil$. Differentiating this equation $n$ times respect to $\lambda$ and then setting $\lambda=0$ gives us the time-ordered cumulant and $\hat{\cal G}^{(n)}(t)$. Doing so yields~\cite{Kampen,Fox74,Fox75,Fox76}:
\begin{align}
\int_0^t\D s\,\hat{\cal G}^{(n)}(s)&=i^n\,\int_0^t\D t_1 \int_0^{t_1}\D t_2\int_0^{t_2}\D t_3\cdots\int_0^{t_{n-1}}\D t_n\,
\mathbb{E}^{(n)}_c\left(
\hat{\cal V}(t_1)\,\hat{\cal V}(t_2)\,\hat{\cal V}(t_3)\cdots\hat{\cal V}(t_n)
\right)
\\&=(i)^n
\sum_{\sum_{l=1}^\infty lm_l=n}\,(-1)^{p-1}\,(p-1)!\,
\left\lceil
\prod_{l=1}^\infty\,\frac{1}{m_l!}
\left(\int_0^t\D s \,\hat{\cal A}^{(l)}(s)\right)^{m_l}
\right\rceil
\end{align}
where $p=\sum_{l=1}^{\infty}m_l$ and
\begin{align}
\hat{\cal A}^{(l)}(t)=\int_0^t\D t_1 \int_0^{t_1}\D t_2\int_0^{t_2}\D t_3\cdots\int_0^{t_{n-1}}\D t_n\,
\mathbb{E}\left(
\hat{\cal V}(t_1)\,\hat{\cal V}(t_2)\,\hat{\cal V}(t_3)\cdots\hat{\cal V}(t_n)
\right)
\end{align}
For example, one has:
\begin{align}
\mathbb{E}^{(1)}_c\left(\hat{\cal V}(t_1)\right)&=\mathbb{E}\left(\hat{\cal V}(t_1)\right)
\\
\mathbb{E}^{(2)}_c\left(\hat{\cal V}(t_1)\hat{\cal V}(t_2)\right)&=\mathbb{E}\left(\hat{\cal V}(t_1)\hat{\cal V}(t_2)\right)
-\mathbb{E}\left(\hat{\cal V}(t_1)\right)\mathbb{E}\left(\hat{\cal V}(t_2)\right)
\\
\mathbb{E}^{(3)}_c\left(\hat{\cal V}(t_1)\hat{\cal V}(t_2)\hat{\cal V}(t_3)\right)&=
\mathbb{E}\left(\hat{\cal V}(t_1)\hat{\cal V}(t_2)\hat{\cal V}(t_3)\right)
-\mathbb{E}\left(\hat{\cal V}(t_1)\right)\mathbb{E}\left(\hat{\cal V}(t_2)\hat{\cal V}(t_3)\right)
-\mathbb{E}\left(\hat{\cal V}(t_1)\hat{\cal V}(t_2)\right)\mathbb{E}\left(\hat{\cal V}(t_3)\right)
\\\nonumber&-\mathbb{E}\left(\hat{\cal V}(t_1)\hat{\cal V}(t_3)\right)\mathbb{E}\left(\hat{\cal V}(t_2)\right)
+\mathbb{E}\left(\hat{\cal V}(t_1)\right)\mathbb{E}\left(\hat{\cal V}(t_2)\right)\mathbb{E}\left(\hat{\cal V}(t_3)\right)
+\mathbb{E}\left(\hat{\cal V}(t_1)\right)\mathbb{E}\left(\hat{\cal V}(t_3)\right)\mathbb{E}\left(\hat{\cal V}(t_2)\right)
\end{align}

The corresponding differential equation of Eq.\eqref{U-cum} is:
\begin{align}\label{dU-cum}
\frac{\D}{\D t}\mathbb{E}\left(\hat{\cal U}^{\text{\tiny I}}(t;0)\right)
=\left(\sum_{n=1}^\infty\,\hat{\cal G}^{(n)}(t)\right)
\,\mathbb{E}\left(\hat{\cal U}^{\text{\tiny I}}(t;0)\right),
\end{align}
which corresponds to the following equation in the Schr\"odinger picture:
\begin{align}\label{dU/dt}
\frac{\D}{\D t}\mathbb{E}\left(\hat{\cal U}(t;0)\right)
=\left(-\frac{\I}{\hbar}\hat{\cal H}+\sum_{n=1}^\infty\,
e^{-\frac{\I}{\hbar}\hat{H}t}\,\hat{\cal G}^{(n)}(t)\,e^{\frac{\I}{\hbar}\hat{H}t}
\right)\,\mathbb{E}\left(\hat{\cal U}(t;0)\right),
\end{align}
where the super-operator $\hat{\cal H}$ is given by $\hat{\cal H}\hat{\rho}=\left[\hat{H},\hat{\rho}\right]$. Remember that at time $t=0$, the stochastic mapping is a non-random term equal to identity. Introducing above equation into Eq.\eqref{U-I} yields:
\begin{align}\label{dRho/dt}
\frac{\D}{\D t}\,\hat{\rho}(t)
=\left(-\frac{\I}{\hbar}\hat{\cal H}+\sum_{n=1}^\infty\,
e^{-\frac{\I}{\hbar}\hat{H}t}\,\hat{\cal G}^{(n)}(t)\,e^{\frac{\I}{\hbar}\hat{H}t}
\right)\,\hat{\rho}(t).
\end{align}
This is the master equation which corresponds to the random Schr\"odinger equation with Hamiltonian $\hat{H}=\hat{H}_0+\hat{V}(t)$ as the one in Eq.(3) in the main text. We now assume the random field is produced by many different independent sources. Therefore we can resort to the central limit theorem~\cite{Kampen} and take $\xi(t,\kk)$ as a Gaussian noise. Without loss of generality, we set the mean value as zero ($\mathbb{E}(\xi(t,\kk))=0$). In the case of a Gaussian noise with zero mean, one finds: $\hat{\cal G}^{(2n-1)}(t)=0$~\cite{Kampen,Fox74,Fox75,Fox76}; however, contrary to ordinary stochastic-averages, $2n$-th orders of time-ordered cumulant stochastic-averages do not vanish for $n\geq 2$; that is to say:
\begin{align}
\mathbb{E}\left(\hat{\cal V}(t_1)\hat{\cal V}(t_2)\cdots\hat{\cal V}(t_{2n})\right)=0;
~~~~~
\mathbb{E}^{(2n)}_c\left(\hat{\cal V}(t_1)\hat{\cal V}(t_2)\cdots\hat{\cal V}(t_{2n})\right)\neq0;
~~~~~\text{for } n\geq 2.
\end{align}
This is an implication of the non-commutativity of $\hat{V}^{\text{\tiny I}}(t)$ at different times. However, if the random field is white in time (i.e. $\mathbb{E}(\xi(t,\kk)\xi(t',\kk'))\propto \delta(t-t')$) then $\mathbb{E}_c\left(\hat{\cal V}(t_1)\hat{\cal V}(t_2)\cdots\hat{\cal V}(t_{2n})\right)=0$ for $n\geq2$. Accordingly, given random field is a Gaussian random field with zero mean, Eq.\eqref{dRho/dt} reduces to:
\begin{align}\label{dRho/dt-2}
\frac{\D}{\D t}\,\hat{\rho}(t)
=\left(-\frac{\I}{\hbar}\hat{\cal H}+\sum_{n=1}^\infty\,
e^{-\frac{\I}{\hbar}\hat{H}t}\,\hat{\cal G}^{(2n)}(t)\,e^{\frac{\I}{\hbar}\hat{H}t}
\right)\,\hat{\rho}(t).
\end{align}
Now if one argues that the most important term in above summation comes from $n=1$, then the evolution of the density matrix is dominated by the second time-ordered cumulant, which is given by:
\begin{align}
\frac{\D}{\D t}\hat{\rho}(t)=&
\left(-\frac{\I}{\hbar}\hat{\cal H}+
e^{-\frac{\I}{\hbar}\hat{H}t}\,\hat{\cal G}^{(2)}(t)\,e^{\frac{\I}{\hbar}\hat{H}t}
\right)\,\hat{\rho}(t)
\\=&-\frac{\I}{\hbar}\com{\hat{H}}{\hat{\rho}(t)}
-\int_0^t\D s\,\mathbb{E}\left(\com{ \hat{V}(t) }{ \com{
e^{\frac{\I}{\hbar}\hat{H}(s-t)}\,\hat{V}(s)e^{-\frac{\I}{\hbar}\hat{H}(s-t)}
}{\hat{\rho}(t)}}\right).
\label{ms-eq00}
\end{align}

\subsection{Switching between Stratonovich and It\^o differential equations}\label{sto-switch}
In this section, we shall briefly discuss how to switch between Stratonovich and It\^o stochastic differential equations. We used Eqs. (10.2.5) to (10.2.7) and the theorem (8.5.5) in~\cite{sto1}. For a simpler version, please see~\cite[p. 99]{sto2}. Consider a linear stochastic differential equation of the following form:
\begin{eqnarray}
\D X_t&=&f(t,X_t)\D t+\mathbf{G}(t,X_t)\D \mathbf{W}_t
\\
&=&\left(A(t)X_t+a(t)\right)\D t+\mathbf{G}(t,X_t)\D \mathbf{W}_t
\end{eqnarray}
where
\begin{eqnarray}
X_t=
\begin{pmatrix}
  X_t^1  \\
  \vdots  \\
  X_t^d 
\end{pmatrix};
&
A(t)=
\begin{pmatrix}
  A_{11}(t) & \cdots & A_{1d}(t)  \\
  \vdots    & \vdots & \vdots \\
  A_{d1}(t) & \cdots & A_{dd}(t) 
\end{pmatrix};
&
a(t)=
\begin{pmatrix}
  a^1(t)  \\
  \vdots  \\
  a^d(t) 
\end{pmatrix},
\end{eqnarray}
\begin{eqnarray}
\mathbf{G}(t,x)&=&\left(B^1(t)x+b_1(t),\cdots,B^m(t)x+b_m(t)\right)
\end{eqnarray}
\begin{eqnarray}\label{B-matrix}
B^i(t)=
\begin{pmatrix}
  B^i_{11}(t) & \cdots & B^i_{1d}(t)  \\
  \vdots    & \vdots & \vdots \\
  B^i_{d1}(t) & \cdots & B^i_{dd}(t) 
\end{pmatrix};
&&
b_i(t)=
\begin{pmatrix}
  b_i^1(t)  \\
  \vdots  \\
  b_i^d(t) 
\end{pmatrix}
\end{eqnarray}
\begin{eqnarray}
\mathbf{W}_t&=&
\begin{pmatrix}
  W_t^1  \\
  \vdots  \\
  W_t^m 
\end{pmatrix}
\end{eqnarray}
Accordingly, one gets:
\begin{eqnarray}
\label{eq:sde}
\D X_t&=&\left(A(t)X_t+a(t)\right)\D t+\sum_{i=1}^m\left(B^i(t)X_t+b_i(t)\right)\D W^i_t
\\
\D 
\begin{pmatrix}
  X_t^1  \\
  \vdots  \\
  X_t^d 
\end{pmatrix}
&=&
\left[
\begin{pmatrix}
  A_{11}(t) & \cdots & A_{1d}(t)  \\
  \vdots    & \vdots & \vdots \\
  A_{d1}(t) & \cdots & A_{dd}(t) 
\end{pmatrix}
\begin{pmatrix}
  X_t^1  \\
  \vdots  \\
  X_t^d 
\end{pmatrix}
+\begin{pmatrix}
  a^1(t)  \\
  \vdots  \\
  a^d(t) 
\end{pmatrix}
\right]\D t
\\\nonumber&+&
\sum_{i=1}^m 
\left[
\begin{pmatrix}
  B^i_{11}(t) & \cdots & B^i_{1d}(t)  \\
  \vdots    & \vdots & \vdots \\
  B^i_{d1}(t) & \cdots & B^i_{dd}(t) 
\end{pmatrix}
\begin{pmatrix}
  X_t^1  \\
  \vdots  \\
  X_t^d 
\end{pmatrix}
+\begin{pmatrix}
  b_i^1(t)  \\
  \vdots  \\
  b_i^d(t) 
\end{pmatrix}
\right]\D W_t^i
\end{eqnarray}

The correspondence between Stratonovich SDE and Ito SDE is as follows:
\begin{eqnarray}
\label{Stratonovich}\text{(Stratonovich)}\quad 
\D X_t&=&f(t,X_t)\D t+\mathbf{G}(t,X_t)\D \mathbf{W}_t
\\\label{Ito}
\text{(It\^o)}\quad
\D X_t&=&
f(t,X_t)\D t+\mathbf{G}(t,X_t)\D \mathbf{W}_t
\\\nonumber&&
+\frac{1}{2}\sum_{j=1}^m\sum_{k=1}^d
\left(\left(b_j^k(t)+\sum_{l=1}^d B_{kl}^j(t)X_t^l\right)
\frac{\partial \left(B^j(t)X_t\right)}{\partial X^k_t}
\right)\D t
\end{eqnarray}
and likewise:
\begin{eqnarray}
\text{(It\^o)}\quad 
\D X_t&=&f(t,X_t)\D t+\mathbf{G}(t,X_t)\D\mathbf{W}_t
\\
\text{(Stratonovich)}\quad
\D X_t&=&
f(t,X_t)\D t+\mathbf{G}(t,X_t)\D\mathbf{W}_t
\\\nonumber&&
-\frac{1}{2}\sum_{j=1}^m\sum_{k=1}^d
\left(\left(b_j^k(t)+\sum_{l=1}^d B_{kl}^j(t)X_t^l\right)
\frac{\partial \left(B^j(t)X_t\right)}{\partial X^k_t}
\right)\D t
\end{eqnarray}
where
\begin{eqnarray}
\frac{\partial \left(B^j(t)X_t\right)}{\partial X^k_t}
&=&
\begin{pmatrix}
  B^j_{1k}(t)  \\
  \vdots  \\
  B^j_{dk}(t) 
\end{pmatrix}
\end{eqnarray}


\begin{thebibliography}{99}

\bibitem{exp_MW}
K. Hornberger, S. Gerlich, P. Haslinger, S. Nimmrichter and M. Arndt, Rev. Mod. Phys. {\bf 84}, 157 (2012).

\bibitem{exp_MW2}
T. Juffmann, H. Ulbricht and M. Arndt, Rep. Prog. Phys. {\bf 76}, 086402 (2013).

\bibitem{exp_MW3}
M. Arndt and K. Hornberger, Nat. Phys. {\bf 10}, 271 (2014).

\bibitem{dirk} W. Marshall, C. Simon, R. Penrose, and D. Bouwmeester, Phys. Rev. Lett. {\bf 91}, 130401 (2003).

\bibitem{GRW} G.C. Ghirardi, A. Rimini, and T. Weber, Phys. Rev. D {\bf 34}, 470 (1986).

\bibitem{CSL} G.C. Ghirardi, P. Pearle, and A. Rimini, Phys. Rev. A {\bf 42}, 78 (1990);
G. C. Ghirardi, R. Grassi, and F. Benatti, Found. Phys. 25, {\bf 5} (1995).

\bibitem{collapse_review1}
A. Bassi, and G.C. Ghirardi, Phys. Rep. {\bf 379}, 257 (2003).

\bibitem{collapse_review2}
A. Bassi, K. Lochan, S. Satin, T. P. Singh, and H. Ulbricht, Rev. Mod. Phys. {\bf 85}, 471 (2013).

\bibitem{adler} S.L. Adler, J. Phys. A {\bf 40}, 2935 (2007).
\bibitem{bassi} M. Bahrami, M. Paternostro, A. Bassi, and H. Ulbricht, Phys. Rev. Lett. {\bf 112}, 210404 (2014).

\bibitem{nimmrichter} S. Nimmrichter, K. Hornberger, and K. Hammerer, Phys. Rev. Lett. {\bf 113}, 020405 (2014).

\bibitem{diosi} L. Diosi, Phys. Rev. Lett. {\bf 114}, 050403 (2015).

\bibitem{goldwater} D. Goldwater, M. Paternostro, P.F. Barker, Phys. Rev. A {\bf 94}, 010104 (2016).

\bibitem{Vinante}
A Vinante, M Bahrami, A Bassi, O Usenko, G Wijts, TH Oosterkamp, Phys. Rev. Lett. {\bf 116} (9), 090402

\bibitem{Vinante2}
A Vinante, R Mezzena, P Falferi, M Carlesso, A Bassi,
Phys. Rev. Lett. {\bf 119} (11), 110401

\bibitem{adler2}
S. L. Adler, {\it Minimum Interior Temperature for Solid Objects Implied by Collapse Models}: arXiv:1712.01071 (2017).

\bibitem{Piscicchia}
K. Piscicchia, {\it et al}, {\it CSL Collapse Model Mapped with the Spontaneous Radiation}: arXiv:1710.01973.

\bibitem{Callaway}
J. Callaway, {\it Quantum Theory of the Solid State}, 2nd ed (Academic Press, San Diego, 1991).

\bibitem{Bah-Juan}
J. Atalaya, M Bahrami, LP Pryadko, AN Korotkov, Phys. Rev. A 95 (3), 032317 (2017).

\bibitem{Cu}
M.E. Straumanis, L. S. Yu, Acta Cryst. A25, 676-682 (1969).

\bibitem{TeO2-primitive}
S. Cebricin {\it et al.}, Astroparticle Physics 10, 397-404 (1999). 

\bibitem{mw}
L. Mandel and E. Wolf, {\it Optical Coherence and Quantum Optics} (Cambridge: Cambridge University Press 1995).

\bibitem{Loudon}
R. Loudon, {\it The Quantum Theory of Light} (Oxford: Clarendon 1985).

\bibitem{gc}
J. C. Garrison and R. Y. Chiao, Quantum Optics (Oxford
University Press, Oxford, UK, 2008).

\bibitem{adler3}
S. L. Adler, {\it Heating Through Phonon Excitation Implied by Collapse Models}, arXiv:1801.00509v3 (2018).

\bibitem{Lev}
L. Kantorovich, {\it Quantum Theory of the Solid State: An Introduction} (Springer Netherlands 2004).

\bibitem{Frank}
T. l. Bergman, L. S. Lavine, E E Incropera and D. P. DeWitt, {\it Fundamentals of Heat Transfer}, 7th ed. (Wiley, New York, 2011). 


\bibitem{NOSV-Cu}
L.Risegari {\it et al}, Cryogenics 44, 167–170 (2004).

\bibitem{CUORE}
C. Alduino {et al.}: {\it The projected background for the CUORE experiment}, arXiv:1704.08970v2 (2017).


\bibitem{Kampen}
N. G. van Kampen, {\it Stochastic Processes in Physics and Chemistry} (North-Holland, Amsterdam 1981); Chapter I section 6 and 7.

\bibitem{Fox74}
R. F. Fox, J. Math. Phys. 15, 1479 (1974). 
\bibitem{Fox75}
R. F. Fox, J. Math. Phys. 16, 289 (1975). 
\bibitem{Fox76}
R. F. Fox, J. Math. Phys. 17, 1148 (1976). 

\bibitem{sto}
R. F. Fox, J. Math. Phys. 13, 1196 (1972); Phys. Rep. 48, 179 (1978).

\bibitem{sto1}
L. Arnold, {\it Stochastic differential equations: Theory and applications} (Wiley, New York 1974).

\bibitem{sto2}
C. W. Gardiner, {\it Handbook of Stochastic Methods: for Physics, Chemistry and the Natural Sciences}, 2nd Ed. (Springer 1996).

\end{thebibliography}
\end{document}